\documentclass[twoside,11pt]{article}

\usepackage[cp1251]{inputenc}

\usepackage{graphicx}

\usepackage{amssymb}
\usepackage{amsmath}
\usepackage{amsfonts}
\usepackage{wasysym}

\usepackage{titlesec}

\usepackage{cite}

\titleformat*{\section}{\bfseries}
\titleformat*{\subsection}{\bfseries}

\begin{document}           % End of preamble and beginning of text.

\title{\Large 
Generalized uncertainty principle in quantum cosmology for the maximally symmetric space}
\author{V.E. Kuzmichev, V.V. Kuzmichev\\[0.5cm]
\itshape Bogolyubov Institute for Theoretical Physics,\\
\itshape National Academy of Sciences of Ukraine, Kiev, 03143 Ukraine}

\date{}

\maketitle

\begin{abstract}
The new uncertainty relation is derived in the context of the canonical quantum theory with gravity for the case of the maximally symmetric space. 
This relation establishes a connection between fluctuations of the quantities which determine the intrinsic and extrinsic curvatures of the spacelike 
hypersurface in spacetime and introduces the uncertainty principle for quantum gravitational systems. The generalized time-energy uncertainty 
relation, which takes into account gravity, is proposed. It is shown that known Unruh's uncertainty relation follows, as a particular case, from the 
new uncertainty relation. As an example, the sizes of fluctuations of the scale factor and its conjugate momentum are calculated within 
an exactly solvable model.  All known modifications of the uncertainty principle deduced previously from different approaches in the theory of gravity 
and string theory are obtained as particular cases of the proposed general expression. 
\end{abstract}

PACS numbers: 98.80.Qc, 98.80.Cq, 95.35.+d, 95.36.+x 

\section{Introduction}\label{sec:1}
The Heisenberg uncertainty principle plays a fundamental role in quantum mechanics. It states that two observables which do not
commute cannot be measured simultaneously with arbitrary accuracy \cite{deB,Mes}. Already a long time ago it was recognized that 
the inclusion of gravitational interaction into the fabric of quantum theory should lead to the modification of the Heisenberg
uncertainty relation \cite{Bro36a,Bro36b,Wig57,And64}.
It is expected that, at scales less than the Planck scale, the classical concepts of space and time lose their meanings and
the radical revision of our notions of them is required. The minimum length of the order of the Planck length appears
as the new ingredient of the theory with gravity and determines a natural restriction on measurable distances.

Possible modifications of the Heisenberg uncertainty relations which take into account effects of gravity have been debated since the middle 1980's.
The modification of the position-momentum uncertainty relation for a test particle moving in a gravitational field 
was formulated and the existence of a Planck-scale minimal observable length was shown 
in string theory \cite{Ama,Kon,Wit}. The same result was obtained using general model-independent properties of a quantum theory of gravitation
\cite{Mag,Gar} and taking into account the constraint on an upper limit of the acceleration of massive particles \cite{Cap}.
The consequences of the existence of a minimal length, the quantum-mechanical structure which underlies it \cite{Kem} and
its effect on the physical properties of various objects (the hydrogen atom spectrum \cite{Bra}, the Lamb shift, Landau levels, and others \cite{Das}) 
were investigated (an extensive bibliography can be found, e.g. in Refs.~\cite{Taw,Ber}). The existence of a minimum observable momentum
can lead to a more general understanding of the influence of gravity on the dynamics of the system \cite{Bam}.

It seems indubitable that for clarification of the influence of spacetime curvature effects on the magnitude of dispersions of two observables
corresponding to canonically conjugate variables (such as position and momentum, time and energy), one should have at its disposal the 
quantum theory which treats gravity on the same grounds as quantized matter fields. A consistent quantum theory of gravity, in principle, 
can be constructed on the basis of the Arnowitt-Deser-Misner (ADM) Hamiltonian formalism \cite{ADM} of general relativity with the application of the 
canonical quantization method. The  canonical approach to quantization, successful in constructing the 
nonrelativistic quantum mechanics and quantum field theories in the flat spacetime, encounters well-known difficulties when applied to
gravity, such as understanding of the time evolution, the divergence of the norm of the state vectors, the measurement problem and others. 
The structure of constraints in general relativity is such that variables which correspond to true 
dynamical degrees of freedom cannot be singled out. This is stipulated by an absence of predetermined way to identify spacetime 
events in generally covariant theory \cite{Kuch}. The quantum theory of gravity based on the Wheeler-DeWitt
equation makes no reference to a time parameter \cite{Whe,DeW}. Attempts to identify one of the matter field
variables of the equations of quantum geometrodynamics with time weren't successful. However, time
can be introduced in the theory in a special manner by bringing in an additional matter source 
(similar to DeWitt's relativistic elastic media with clocks) in the form of a perfect (reference) fluid which defines a dynamical reference system 
(material reference frame) \cite{Ish,Bro}. 

A model with a finite number of degrees of freedom may provide a reasonable framework for addressing the problems of quantum gravity. 
The homogeneous minisuperspace models have been proven to be successful -- consistent with observations and having predictive power -- in 
classical cosmology. This appears explicable, in view of the fact that the Universe can, to first approximation, be considered as being homogeneous, 
and gives rise to the hope that homogeneous models could be useful in quantum cosmology as well.
For such models, the quantum theory of gravity with a well-defined time variable was proposed and studied in Refs. \cite{Kuz08,Kuz08a,Kuz13,Kuz16,Kuz17}.

In the present paper, we study the fluctuations of the observables that characterize the quantum gravitational system, like the
intrinsic and extrinsic curvatures of the spacelike hypersurface in spacetime.
In order to formulate the uncertainty relation, we require well-defined state vectors which allow us to calculate the expectation values of observables
and their statistical fluctuations. In Sect.~2, the version of quantum theory of gravitational system in the maximally symmetric space is given.
The distinguished feature of this theory is that one quantizes the observable represented by some Hermitian operator, which plays the role of the 
effective Hamiltonian and is measured in units of the conversion constant ($\hbar c$). Therefore, difficulties with definition of energy in general 
relativity and in corresponding quantum field theory constructed on its basis do not arise in this approach. Sect.~3 is devoted to the derivation
of the uncertainty principle for quantum systems with gravitational interaction. It is shown that the Heisenberg uncertainty relations
can be reformulated in terms of geometrical values such as intrinsic and extrinsic curvatures. In Sect.~4, the new
generalized uncertainty relation of energy and time is formulated and its connection with the geometrical variables of the theory is shown.
The goal of Sect.~5 is to reduce the obtained uncertainty relation for quantum gravitational systems to the Unruh's relation between the
metric and the curvature. In Sect.~6, the fluctuations  of cosmological parameters are calculated within an exactly solvable model.
 In Sect.~7, the effects of gravity are extracted explicitly and
the uncertainty relation for fluctuations of position and momentum of a test particle, which takes into account gravitational interaction,
are obtained. From this relation, it follows the generalized uncertainty principle
proposed previously by a number of authors by using different quantum theories with gravity. The consequences of the existence of
minimum length and minimum momentum of a test particle which can be measured are also discussed here. 
Finally, conclusions are drawn in Sect.~8.

\section{Quantization scheme}\label{sec:2}
Consider the homogeneous isotropic quantum gravitational system (QGS).
In the case of the maximally symmetric geometry with the Robertson-Walker metric, the geometrical properties of the system
are determined by a single variable, namely the cosmic scale factor $a$. The matter sector of the QGS is taken
in the form of a uniform scalar field $\phi$ with a self-adjoint Hamiltonian $H_{\phi}$. This Hamiltonian is defined in a curved spacetime
and it depends on a scale factor $a$ as a parameter, $H_{\phi} = H_{\phi} (a)$. The scalar field can be interpreted as a surrogate of 
all possible real physical fields of matter averaged with respect to spin, space and other degrees of freedom. 
In addition, it is accepted that QGS is filled with a perfect fluid in the form of a relativistic matter with the energy density 
$\rho_{\gamma} = E / a^{4}$, where $E$ is constant. A perfect fluid coupled to a fleet of clocks can be employed as a material reference frame 
enabling one to recognize the instants of time. Following Dirac's approach to quantum gravity \cite{Dir58}, we do not solve constraints prior to
quantization, but convert the second-class constraints into the first-class ones which become constraints
on the state vector (wave function)  $\langle a, \phi | \Psi (T) \rangle$ in the representation of the gravitational and matter fields, $a$ and $\phi$,
the parameter $T$ is a conformal time. The basic equations of such a QGS appear as the set of two differential equations in partial derivatives \cite{Kuz08,Kuz08a,Kuz13,Kuz16,Kuz17} \footnote{Here, for simplicity, the rescaling of the variable $T$ is done and the corresponding
coefficient is included into the definition of proper time.},
\begin{equation}\label{1}
\left(-i \partial_{T} - \frac{1}{2} E \right)| \Psi (T) \rangle = 0,
\quad
\left(- \partial_{a}^{2} + \kappa a^{2} - 2 a H_{\phi} - a^{4} \frac{\Lambda}{3} - E \right)| \Psi (T) \rangle = 0,
\end{equation}
where the curvature constant $\kappa = +1,0,-1$ for spatially closed, flat, and open QGS, respectively, and $\Lambda$ is a cosmological 
constant. The conformal time $T$ is expressed in radians, the cosmic scale factor $a$ is measured in units of Planck length 
$l_{P} = \sqrt{2 G \hbar / (3 \pi c^{3})}$, $G$ is Newton's gravitational constant, the Hamiltonian $H_{\phi}$ is taken in the units of Planck energy 
$m_{P} c^{2} = \hbar c / l_{P}$, the energy density $\rho_{\gamma}$ is in the units of Planck density $\rho_{P} = 3 c^{4} / (8 \pi G l_{P}^{2})$,
so that $E$, which in ordinary physical units has the dimension of [Energy $\times$ Length] $ = [\hbar c]$, becomes dimensionless. In such units, the
commutation relation between $a$ and its conjugate momentum $\pi = - i \hbar\, \partial_{a}$ takes the form
\begin{equation}\label{2}
[a,-i \partial_{a}] = i.
\end{equation}

The equations (\ref{1}) can be combined into one single Schr\"{o}dinger-type time equation
\begin{equation}\label{3}
- i \partial_{T} | \Psi (T) \rangle = \mathrm{H} | \Psi (T) \rangle,
\end{equation}
where the operator 
\begin{equation}\label{4}
\mathrm{H} = \frac{1}{2} \left(- \partial_{a}^{2} + \kappa a^{2} - 2 a H_{\phi} - a^{4} \frac{\Lambda}{3} \right)
\end{equation}
can be considered as the effective Hamiltonian of the QGS which does not depend on time $T$ explicitly and the minus sign is a 
consequence of the gravitational field equations in general relativity. From the condition of self-adjointness of 
$H_{\phi}$ and reality of $a$, it follows self-adjointness of $\mathrm{H}$.

By defining the unitary evolution operator
\begin{equation}\label{5}
U(T,T_{0}) = e^{i \mathrm{H} (T - T_{0})}, 
\end{equation}
where $T_{0}$ is an arbitrary constant taken as a time reference point, the solution of Eq.~(\ref{3}) can be written as follows
\begin{equation}\label{6}
| \Psi (T) \rangle = U(T,T_{0}) | \Psi_{H} (T_{0}) \rangle,
\end{equation}
where $| \Psi_{H} (T_{0}) \rangle$ is the state vector in the Heisenberg representation.

The second equation from the set (\ref{1}) which defines the change of the state vector $| \Psi (T) \rangle$ as a function of $a$ and $\phi$ 
can be integrated with respect to $\phi$. With this purpose
we introduce the complete set of orthonormalized functions $\langle \chi | u_{k} \rangle$ in a representation of a rescaled variable
$\chi = \chi (a, \phi)$, in which the Hamiltonian $H_{\phi}$ is diagonalized,
\begin{equation}\label{7}
\langle u_{k} | H_{\phi} | u_{k'} \rangle = M_{k}(a)\,\delta_{k k'},
\end{equation}
where $M_{k}(a)$ is the proper mass-energy of the new effective matter in discrete and/or continuous $k$th state obtained after averaging 
of $H_{\phi}$ with respect to the field $\chi$ in a comoving volume $\frac{1}{2} a^{3}$. 
For example, in the case, when $H_{\phi}$ describes the homogeneous scalar field, 
the new effective matter is a barotropic fluid with the energy density $\rho_{m} = 2 M_{k}(a) / a^{3}$ and the pressure 
$p_{m} = - \frac{1}{3} (d \ln M_{k}(a) / d \ln a) \rho_{m}$ \cite{Kuz16}.

We, further, bring in another complete set of orthonormalized functions $\langle a | f_{nk} \rangle$, 
which satisfies the equation
\begin{equation}\label{8}
\left(- \partial_{a}^{2} + \kappa a^{2} - 2 a M_{k}(a) - a^{4} \frac{\Lambda}{3} \right) | f_{nk} \rangle = E_{n} | f_{nk} \rangle,
\end{equation}
where $n$ enumerates discrete and/or continuous states of the QGS with matter in fixed $k$th state. Then, the vector $| \Psi_{H} (T_{0}) \rangle$
can be written as the superposition
\begin{equation}\label{9}
| \Psi_{H} (T_{0}) \rangle = \sum_{n, k} C_{n k} (T_{0}) | u_{k} \rangle | f_{nk} \rangle,
\end{equation}
where the coefficient $C_{n k} (T_{0})$ gives the probability $|C_{n k} (T_{0})|^{2}$ to find
the QGS in $n$th state of relativistic matter and $k$th state of averaged effective matter at the instant of time $T_{0}$.
Since, by definition, the vectors $| u_{k} \rangle$ and $| f_{nk} \rangle$ exhaust all the possible states of the matter components of the QGS,
the normalization condition for $| \Psi (T) \rangle$ takes the form
\begin{equation}\label{10}
\langle \Psi (T) | \Psi (T) \rangle = \langle \Psi_{H} (T_{0}) | \Psi_{H} (T_{0}) \rangle = \sum_{n, k} |C_{n k} (T_{0})|^{2} = 1.
\end{equation}
The transition amplitude $T_{0} \rightarrow T$ determines
the mean value of the evolution operator with respect to the state (\ref{9}),
\begin{equation}\label{11}
\langle \Psi (T_{0}) | \Psi (T) \rangle =
\langle \Psi_{H} (T_{0}) | U(T,T_{0})  |\Psi_{H} (T_{0}) \rangle = \sum_{n, k} e^{\frac{i}{2} E_{n} (T - T_{0})} |C_{n k} (T_{0})|^{2}.
\end{equation}
It demonstrates that the evolution of the system in time $T$ has a periodic character in each $n$th state. Taking into account 
self-agjointness of the effective Hamiltonian (\ref{4}), from Eq.~(\ref{3}) and its complex conjugate, it follows the equation
\begin{equation}\label{12}
\frac{d \langle \mathrm{A} \rangle }{d T} = \frac{1}{i} \langle [\mathrm{H},\mathrm{A}] \rangle + \langle \frac{\partial \mathrm{A}}{\partial T} \rangle,
\end{equation}
which gives the time-dependence of the mean value of an observable $\mathrm{A}$,
\begin{equation}\label{13}
\langle \mathrm{A} \rangle \equiv \langle \Psi (T) | \mathrm{A} | \Psi (T) \rangle = \langle \Psi_{H} (T_{0}) | \mathrm{A}_{H} | \Psi_{H} (T_{0}) \rangle
\equiv \langle \mathrm{A}_{H} \rangle_{H},
\end{equation}
where $\mathrm{A}_{H} = U^{\dagger} \mathrm{A} U$ is the observable of the Heisenberg representation. From the unitary property of 
the operator $U$ (\ref{5}), we have
\begin{equation}\label{14}
\langle [\mathrm{H},\mathrm{A}] \rangle = \langle [\mathrm{H}_{H},\mathrm{A}_{H}] \rangle_{H},
\end{equation}
where $\mathrm{H}_{H} = U^{\dagger} \mathrm{H} U$ is the Hamiltonian of the Heisenberg representation.

Let us note that, according to Eq.~(\ref{9}), the equations (\ref{7}) and (\ref{8}) determine the stationary quantum states of the QGS
at some fixed instant of time $T_{0}$, the choice of which is arbitrary, $\langle \chi | u_{k} \rangle \equiv u_{k} (\chi, T_{0})$,
and $\langle a | f_{nk} \rangle \equiv f_{nk} (a, T_{0})$.

\section{Uncertainty principle}\label{sec:3}
The uncertainty relation for two observables in a Hilbert space, $\mathrm{A}$ and $\mathrm{B}$, which do not depend upon the time $T$ explicitly
can be written as (cf. Ref.~\cite{Mes})
\begin{equation}\label{15}
\Delta \mathrm{A} \Delta \mathrm{B} \geq \frac{1}{2} |\langle [\mathrm{A},\mathrm{B}] \rangle|,
\end{equation}
where $\Delta \mathrm{A} = \sqrt{\langle \mathrm{A}^{2} \rangle - \langle \mathrm{A} \rangle^{2}}$, 
$\Delta \mathrm{B} = \sqrt{\langle \mathrm{B}^{2} \rangle - \langle \mathrm{B} \rangle^{2}}$ are
the root-mean-square deviations of $\mathrm{A}$ and of $\mathrm{B}$, respectively.

Let $\mathrm{A} = a$ and $\mathrm{B} = \pi = -i \partial_{a}$. Then, taking into account Eq.~(\ref{2}), we find
the uncertainty relation between the scale factor and its conjugate momentum
\begin{equation}\label{16}
\Delta a \Delta \pi \geq \frac{\hbar}{2}
\end{equation}
(in ordinary physical units). This relation coincides in form with the uncertainty relation between position and momentum in ordinary
quantum mechanics, but it has a different physical meaning. It is expressed in geometrical quantities and thus describes the effects of 
spacetime curvature. The inequality (\ref{16}) reduces to an equality, if $\Delta a = l_{P}$ and $\Delta \pi = \frac{1}{2} m_{P} c$. 
It can be considered as an indication of the existence of a minimum length equal to the Planck length. 

The uncertainty relation (\ref{16}) establishes, in fact,
a connection between fluctuations of the quantities which determine the intrinsic and extrinsic curvatures of the spacelike hypersurface
in spacetime. Using the relations for the scalar curvature $^{(3)}R$ and the extrinsic curvature tensor
$K_{ij} = -\frac{1}{2} \partial ^{(3)}g_{ij} / \partial \tau$, where $^{(3)}g_{ij}$ is the 3-metric and $\tau$ is the proper time,
Eq.~(\ref{16}) can be written explicitly in terms of curvature fluctuations,
\begin{equation}\label{17}
\Delta ^{(3)}R \ \Delta K \gtrsim 4 \pi \hbar \ \frac{ |^{(3)}R|}{^{(3)}V},
\end{equation}
where $K = K_{i}^{i}$ and $^{(3)}V \sim \frac{4}{3} \pi a^{3}$ is the 3-volume of the measurement (observed part of the QGS).

\section{Generalized time-energy uncertainty relation}\label{sec:4}
Let us find the generalized time-energy uncertainty relation, which takes into account gravity. 
Let $\mathrm{A}$ be an observable which does not depend on time explicitly and does not commute with the effective Hamiltonian (\ref{4}).
Setting $\mathrm{B} = \mathrm{H}$ in Eq.~(\ref{15}), bearing in mind Eq.~(\ref{12}) and that the proper time $\tau$ is
connected with the conformal time $T$ by the differential equation $c d\tau = a\, dT$,
we obtain the uncertainty relation in ordinary physical units
\begin{equation}\label{18}
\tau_{\mathrm{A}}\, \frac{\Delta \mathrm{E}}{|a|} \geq \frac{\hbar}{2},
\end{equation}
where $\Delta \mathrm{E}$ is the root-mean-square deviation of $\mathrm{H}$, and
\begin{equation}\label{19}
\tau_{\mathrm{A}} = \left|\Delta \mathrm{A}\,\left(\frac{d \langle \mathrm{A} \rangle}{d \tau} \right)^{-1} \right|
\end{equation}
is a time characteristic of the evolution of the statistical distribution of $\mathrm{A}$ (i.e. the time necessary for this statistical distribution to be
considerably modified).  The quantity $\Delta \mathrm{E} / |a|$ is the statistical fluctuation of the result of the energy measurement,
where the denominator takes into account the redshift correction due to the expansion of the QGS. 
In the limit $|a| \rightarrow 0$, this fluctuation becomes infinitely large, while a time characteristic $\tau_{\mathrm{A}}$ can acquire any value 
in accordance with the uncertainty relation (\ref{18}). Thus, near the initial cosmological singularity notions of time and energy lose their meaning 
(they cannot be measured). 

Because of the presence of the
multiplier $1 / |a|$, the uncertainty relation (\ref{18}) does not reduce to the time-energy uncertainty relation of ordinary quantum mechanics.
In our approach, the effective Hamiltonian has the dimension of the constant $E$ which is quantized according to Eq.~(\ref{8}), the latter determines
the quantum state of relativistic matter.

For $\mathrm{A} = a$, the relation (\ref{18}) reduces to the uncertainty relation (\ref{16}), where 
$\Delta \pi = \frac{\Delta \mathrm{E}}{c}\left(\frac{d \langle a \rangle}{dT} \right)^{-1}$ is the statistical fluctuation of the momentum.

\section{The Unruh's uncertainty relation}\label{sec:5}
The uncertainty relation (\ref{16}) can be reduced to the Unruh's uncertainty relation between the metric and the curvature.
One may assume formally that Einstein's equations are valid in the quantum regime as well \cite{Unr}. Indeed, as is shown in Ref.~\cite{Kuz13},
the equation (\ref{12}) for $\mathrm{A} = a$ and $\mathrm{A} = \pi = -i \partial_{a}$ can be reduced to the Einstein-Friedmann equations
which contain the quantum correction terms to the total energy density and pressure. Then the rate of change of the momentum in time is
given by the equation $\dot{\pi} = - \frac{1}{2} a^{2} T_{\alpha}^{\alpha} + \kappa$, where $T_{\alpha}^{\alpha}$ is the trace of the stress tensor.
We shall restrict ourselves to the study of the fluctuations of the quantities in spatial directions.
In a comoving reference frame, one can express the $T_{x}^{x}$ component of the stress 
tensor as $T_{x}^{x} = - \mathrm{p}$, where $\mathrm{p}$ is the pressure defined as the force acting on the surface element having an area of $A$
in the direction of $x$. In that case the fluctuation of the momentum can be estimated as $\Delta \pi \sim \Delta T_{x}^{x} A \delta \tau$, where 
$\delta \tau$ is a time interval, $A \sim a^{2}$ and $\Delta T_{x}^{x} \sim T_{x}^{x}$. The metric component $g_{xx}$ can be
represented in the form $g_{xx} = a^{2} \gamma_{xx}$, where $\gamma_{xx}$ is the comoving spatial metric component 
whose fluctuation can be neglected, $\Delta \gamma_{xx} = 0$. Then the fluctuations $\Delta g_{xx}$ and $\Delta a$ are
connected between themselves: $\Delta g_{xx} / g_{xx} = 2 \Delta a / a$. As a result, in the rest frame the relation (\ref{16}) takes the
form
\begin{equation}\label{20}
\Delta g_{xx} \Delta T_{x}^{x} \gtrsim \hbar \frac{g_{xx}}{\delta \tau ^{(3)}V},
\end{equation}
where $^{(3)}V \sim a^{3}$ is the 3-volume. Introducing the Einstein tensor, 
$G_{x}^{x} = 8 \pi T_{x}^{x}$ (in units $G = c = 1$), and defining the 4-volume
$^{(4)}V \sim \delta \tau^{(3)}V$, we rewrite the preceding relation in Unruh's form
\begin{equation}\label{21}
\Delta g_{xx} \Delta G^{xx} \gtrsim \hbar \frac{8 \pi}{^{(4)}V}.
\end{equation}
From Eq.~(\ref{21}) the Heisenberg position-momentum uncertainty relation which takes into account the effects of gravity on quantum fields 
can be restored (cf. Ref.~\cite{Unr}). 

Equation (\ref{20}) can also be represented as the relation
\begin{equation}\label{22}
\delta \varepsilon\, \delta \tau \gtrsim  \hbar,
\end{equation}
which connects the deviation of energy $\delta \varepsilon = \Delta g_{xx} \Delta T^{xx}\, ^{(3)}V$ with the time interval $\delta \tau$
(cf. with Eq.~(\ref{18})).

\section{Exactly solvable model}\label{sec:6}
The uncertainty relation (\ref{16}) and its consequences (\ref{17})-(\ref{22}) determine restrictions on simultaneous measurement of 
the corresponding cosmological parameters of the QGS described by the state vector $| \Psi (T) \rangle$. According to Eqs.~(\ref{6})
and (\ref{9}), such a QGS is a superposition of subsystems (or universes in a multiverse model) characterized by the quantum numbers
$k$ and $n$, i.e., they are filled with the effective matter in $k$th state with definite mass-energy $M_{k}(a)$ and relativistic matter
in $n$th state with the energy $E_{n}/(2 a)$. Each subsystem is described by its wave function $\langle a | f_{nk} \rangle$, 
which satisfies the stationary equation (\ref{8}). The time equation for $| f_{nk} (T) \rangle = e^{\frac{i}{2} E_{n} (T - T_{0})} | f_{nk} \rangle$ 
follows from Eqs.~(\ref{3}), (\ref{6}) and (\ref{9}),
\begin{equation}\label{23}
- i \partial_{T} | f_{nk} (T) \rangle = \mathrm{H}_{k} | f_{nk} (T) \rangle,
\end{equation}
with the ``Hamiltonian'' 
\begin{equation}\label{24}
\mathrm{H}_{k} = \frac{1}{2} \left(- \partial_{a}^{2} + \kappa a^{2} - 2 a M_{k}(a) - a^{4} \frac{\Lambda}{3} \right)
\end{equation}
which is the operator (\ref{4}) averaged over the states $| u_{k} \rangle$, $\mathrm{H}_{k} = \langle u_{k} | \mathrm{H} | u_{k} \rangle$.

The uncertainty relation for the observables $a$ and $\pi = -i \partial_{a}$ in such a subsystem takes the form
\begin{equation}\label{25}
\delta a \, \delta \pi \geq \frac{1}{2}
\end{equation}
(in dimensionless units), where $\delta a = \sqrt{\langle a^{2} \rangle_{n k} - \langle a \rangle^{2}_{n k}}$
and $\delta \pi = \sqrt{\langle \pi^{2} \rangle_{n k} - \langle \pi \rangle^{2}_{n k}}$
are the root-mean-square deviations, and averaging is performed over the $f_{nk}$-states,
e.g. $\langle a^{2} \rangle_{n k} \equiv \langle f_{nk} | a^{2} | f_{nk} \rangle$ and so on.

Let us calculate the fluctuations $\delta a$ and $\delta \pi$ in an explicit form for a specific spatially closed subsystem with zero cosmological 
constant ($\Lambda = 0$) filled with separate (non-interacting) macroscopic bodies (dust) and radiation. The mass of the dust is 
$M_{k}(a) = \mu \left(k + \frac{1}{2} \right) = const$, where $\mu$ is the mass of a single macroscopic body and $k$ is the number of such bodies.
In this case, the solution of Eq.~(\ref{8}) is
\begin{equation}\label{26}
| f_{nk} \rangle \equiv f_{n} (\xi_{k}) = N_{nk}\, e^{- \frac{1}{2} \xi_{k}^{2}}\, H_{n} (\xi_{k}), \quad E_{n} = 2 n + 1 - M^{2}_{k},
\end{equation}
where $\xi_{k} = a - M_{k}$, $H_{n}$ is the Hermite polynomial, $N_{nk}$ is the normalizing constant, and the wave function is normalized 
on the interval $[-M_{k}, \infty )$.

Then, with an accuracy of order $e^{-M^{2}_{k}}$ \cite{Kuz08}, we have
\begin{equation}\label{27}
\langle a^{2} \rangle_{n k} = n + \frac{1}{2} + M^{2}_{k}, \quad \langle a \rangle_{n k} = M_{k},
\end{equation}
so that (in ordinary physical units)
\begin{equation}\label{28}
\delta a = l_{P} \sqrt{n + \frac{1}{2}}.
\end{equation}
For the momentum, we obtain
\begin{equation}\label{29}
\langle \pi^{2} \rangle_{n k} = n + \frac{1}{2}, \quad \langle \pi \rangle_{n k} = 0
\end{equation}
and
\begin{equation}\label{30}
\delta \pi =  m_{P}\, c\, \sqrt{n + \frac{1}{2}}.
\end{equation}
As a consequence, we get the uncertainty product of the same form as for harmonic oscillator,
\begin{equation}\label{31}
\delta a \, \delta \pi = \left(n + \frac{1}{2}\right) \hbar.
\end{equation}
From Eqs.~(\ref{28}) and (\ref{30}), one can see that the fluctuations $\delta a$ and $\delta \pi$ take minimum values in a ground (vacuum) state
with $n = 0$,
\begin{equation}\label{32}
\delta a_{min} = \frac{l_{P}}{\sqrt{2}}, \quad \delta \pi_{min} =  \frac{m_{P}\, c}{\sqrt{2}}.
\end{equation}

\section{Effects of gravity}\label{sec:7}
We will consider an influence of gravity on quantum fluctuations of position and momentum of a test particle.
By introducing the root-mean-square deviations of position $\Delta x$ and momentum $\Delta p$ of motion of a test particle 
in coordinate space, one can rewrite the uncertainty relation (\ref{16}) in an equivalent form, in which the effects of gravity 
may be extracted explicitly on the right-hand side,
\begin{equation}\label{123}
\Delta x \frac{\Delta p}{\hbar} \geq \frac{1}{2} \frac{\Delta x}{\Delta a} \frac{\Delta p}{\Delta \pi}.
\end{equation}
In the zero approximation, when the effects of gravity are considered as negligible compared with the other forces, 
the following conditions have to be satisfied, $\Delta a = \Delta x$, $\Delta \pi = \Delta p$, and Eq.~(\ref{123}) turns into
the Heisenberg uncertainty relation. Those conditions are modified by the gravitational interaction.

Having this in mind, we suppose that the deviations $\Delta a$ and $\Delta \pi$ can be represented as follows,
\begin{equation}\label{124}
\Delta a = \langle \rho(Q) \rangle \Delta x, \quad \Delta \pi =  \langle \zeta (P) \rangle \Delta p,
\end{equation}
where $\rho(Q)$ and $\zeta (P)$ are some functions on deviations of position $Q = x - \langle x \rangle$ and momentum
$P = p - \langle p \rangle$, and averaging ensures independence of $\Delta a$ and $\Delta \pi$ from direction in coordinate space.

Let us define the functions $\rho(Q)$ and $\zeta (P)$ in the form of the normal frequency functions
\begin{equation}\label{125}
\rho(Q) = e^{-\frac{Q^{2}}{2 L_{x}^{2}}}, \quad \zeta (P) = e^{-\frac{L_{p}^{2}}{2} \left(\frac{P}{\hbar}\right)^{2}},
\end{equation}
normalized to satisfy $\rho(0) = 1$, $\zeta (0) = 1$. Here $L_{x}^{2}$ is the dispersion in coordinate space and 
$\left(\frac{\hbar}{L_{p}}\right)^{2}$ is the dispersion in momentum space.

The normal distributions (\ref{125}) correspond to the wave packet representing a test particle localized in coordinate space about 
$Q = 0$ and localized in momentum space about $P = 0$. The quantities  $L_{x}$ and $L_{p}$ have the dimensions of length.
They are independent free parameters, since the normal frequency function $\zeta (P)$ is not a Fourier transform of $\rho(Q)$.
The physical content of the parameters $L_{x}$ and $L_{p}$ is different. As is obvious already from Eqs.~(\ref{123})-(\ref{125}), 
if the fluctuations $\Delta x$ and $\Delta p$ are non-zero, the Heisenberg uncertainty relation
can be restored only in the formal limits $L_{x} \rightarrow \infty$ and $L_{p} \rightarrow 0$ reached simultaneously. If $L_{x} = \infty$, but 
$L_{p} \neq 0$, then gravity contributes to the fluctuation of momentum. In the other case, if $L_{x} < \infty$, but $L_{p} = 0$ 
then gravity affects the fluctuation of position.

The reason why the parameters $L_{x}$ and $L_{p}$ have different physical contents is that position $x$ and momentum $p$ fluctuate 
independently under the action of gravity. Spacetime and momentum space are both dynamical and fluctuating, and momentum
space is independent of spacetime and it cannot be just a Fourier transform of coordinate space (cf. Refs.~\cite{Ame,Cha}).

The uncertainty relation (\ref{123}) with the deviations $\Delta a$ and $\Delta \pi$ (\ref{124}) and
the normal frequency functions (\ref{125}) transforms into the generalized Heisenberg-type uncertainty relation with corrections to gravity,
in the case
\begin{equation}\label{126}
\left(\frac{Q}{L_{x}}\right)^{2} < 1 \quad \mbox{and} \ \left(\frac{L_{p} P}{\hbar}\right)^{2} < 1.
\end{equation}
Taking this into account, let us find the form of corrections to gravity. With this end in view, substituting Eq.~(\ref{124}) and Eq.~(\ref{125}) 
into Eq.~(\ref{123}), expanding the exponentials in Eq.~(\ref{125}) in power series, and truncating after the first corrections to unity,
we obtain the approximate inequality
\begin{equation}\label{127}
\Delta x \frac{\Delta p}{\hbar} \geq \frac{1}{2} \left[1 +  \frac{1}{2 L_{x}^{2}} (\Delta x)^{2} + \frac{L_{p}^{2}}{2} \left(\frac{\Delta p}{\hbar} 
\right)^{2} + \left(\frac{L_{p}}{2 L_{x}}\right)^{2} (\Delta x)^{2} \left(\frac{\Delta p}{\hbar} \right)^{2} + \ldots \right].
\end{equation}
The uncertainty relation (\ref{127}) is invariant under the Balmi-Urban transformations proposed in 
Ref.~\cite{Bam},
\begin{equation}\label{128}
\Delta x \rightarrow \left(L_{x} L_{p} \right) \frac{\Delta p}{\hbar}, \quad
\frac{\Delta p}{\hbar} \rightarrow \left(L_{x} L_{p} \right)^{-1} \Delta x.
\end{equation}
For the future convenience, we rewrite Eq.~(\ref{127}) in the form
\begin{equation}\label{129}
\Delta x \geq \frac{1}{2} \left[ \left(1 +  \frac{1}{2 L_{x}^{2}} (\Delta x)^{2} \right) \frac{\hbar}{\Delta p} +  \frac{L_{p}^{2}}{2}
\left(1 +  \frac{1}{2 L_{x}^{2}} (\Delta x)^{2} \right) \frac{\Delta p}{\hbar} + \ldots \right].
\end{equation}
Following Ref.~\cite{Bam}, we take that $L_{p} = \sqrt{\frac{G \hbar}{c^{3}}}$ is the Planck length and $L_{x} = \sqrt{\frac{3}{\Lambda}}$
is the de Sitter horizon, where $\Lambda$ is a cosmological constant. Then the right-hand side of the inequality (\ref{129}) 
will contain only the fundamental constants
\begin{equation}\label{130}
\Delta x \geq \frac{1}{2} \left[ \left(1 +  \frac{\Lambda}{6} (\Delta x)^{2} \right) \frac{\hbar}{\Delta p} +  \frac{G}{2 c^{3}}
\left(1 +  \frac{\Lambda}{6} (\Delta x)^{2} \right) \Delta p \right].
\end{equation}
From here, under the assumption $\Lambda = 0$, it follows the result obtained in Refs.~\cite{Mag,Cap}
\begin{equation}\label{131}
\Delta x \geq \frac{\hbar}{2 \Delta p} +  \frac{G}{2 c^{3}} \Delta p.
\end{equation}
This relation can be written in the form arising from string theory \cite{Kon,Wit}
\begin{equation}\label{132}
\Delta x \geq \frac{\hbar}{2 \Delta p} +  \frac{\alpha'}{2} \frac{\Delta p}{\hbar},
\end{equation}
where $\alpha' = L_{p}^{2}$ is a fundamental constant controlling the tension of the string. The minimum length in such a theory is equal to
$(\Delta x)_{min} = \sqrt{\alpha'}$ and coincides with the Planck length $\sim 10^{-33}$ cm. From the more general relation (\ref{127}) 
and the Balmi-Urban transformations (\ref{128}), it follows that there should exist not only the minimum length 
$(\Delta x)_{min} = L_{p}$, but also the minimum momentum
$(\Delta p / \hbar)_{min} = 1 / L_{x}$. If $L_{x}$ is the de Sitter horizon, then $(\Delta p / \hbar)_{min} \sim \sqrt{\Lambda} \lesssim 10^{-28}$ 
cm$^{-1}$ for the present-day values of the cosmological parameters. 

By taking the minimum length together with the minimum momentum, one can write the following relation:
$(\Delta x)_{min} (\Delta p / \hbar)_{min} = L_{p} / L_{x} \lesssim 10^{-61}$. Using this relation, it is possible to restore 
the physical parameters $A_{\mbox{\scriptsize today}}$ of the observed part of our Universe from the Planck values
$A_{\mbox{\scriptsize Planck}}$, 
$A_{\mbox{\scriptsize today}} = A_{\mbox{\scriptsize Planck}} (\Delta x)_{min}^{-1} (\Delta p / \hbar)_{min}^{-1}$.
Substituting the Planck length, Planck mass, and Planck time into this expression, we obtain the estimations for
the size of the observed part of the Universe $R_{0} \gtrsim 10^{28}$ cm, for the mass $M_{0} \gtrsim 10^{80}$ GeV, and for the
age of the Universe $t_{0} \gtrsim 10^{17}$ s.

\section{Conclusion}\label{sec:8}
In this paper, it is shown that the Heisenberg-type uncertainty principle can be formulated for variables of the system with gravitational interaction 
evolving in the conformal time. In the case of the maximally symmetric space, the scale factor and its conjugate momentum play
the role of these variables. The obtained uncertainty principle in the form of Eq.~(\ref{16}) can be reformulated through the quantities
which determine the intrinsic and extrinsic curvatures of the spacelike hypersurface in spacetime. As is shown in Eq.~(\ref{17}),
the product of fluctuations of intrinsic and extrinsic curvatures must be greater than or of order the intrinsic curvature per unit volume of the 
measurement.

We draw the attention to the fact that the generalized time-energy uncertainty relation (\ref{18}) contains 
the statistical fluctuation of the result of the measurement of the energy of the relativistic matter which takes into account the correction 
due to the expansion of the system.

Under the assumption that the Einstein equations (with quantum correction terms)
are valid in the quantum regime, we demonstrate that the fundamental relation (\ref{16}) can be reduced to the uncertainty relation
in Unruh's form (\ref{21}). Such a connection between the equations (\ref{16}) and (\ref{21}) may be interpreted as clarifying the physical meaning
of Eq.~(\ref{21}).

Equations (\ref{3}) - (\ref{9}) describe the QGS as a linear superposition of simpler subsystems, each of which is characterized by matter-energy
in specific states with quantum numbers $k$ and $n$. Each separate subsystem is determined by the wave function $\langle a | f_{nk} \rangle$
satisfying Eq.~(\ref{8}). If the subsystem is spatially closed and filled with dust and relativistic matter, then, in the case of zero cosmological 
constant, the equation (\ref{8}) has an analytical solution (\ref{26}) in the form of an oscillator shifted in $a$ axis by the amount of dust mass.
The fluctuations of the scale factor $\delta a$ and its conjugate momentum $\delta \pi$ are quantized according to Eqs.~(\ref{28}) and (\ref{30}), 
while their product (\ref{31}) satisfies the uncertainty relation (\ref{25}), where the equality is reached only for the vacuum state of the subsystem.
In the vacuum state, the fluctuation $\delta a$ has the minimum value of the order of Planck length $l_{P}$ and the fluctuation $\delta \pi$
acquires the minimum value of the order of Planck momentum $m_{P}\, c$ (see Eq.~(\ref{32})).

The size of fluctuations increases as the square root $\sqrt{n}$. It is interesting to estimate the size of fluctuations in the subsystem having 
the mass-energy of the observable part of our universe $l \sim 10^{28}$ cm. The cosmological parameters 
$E_{n} \sim 10^{118} \ll M^{2}_{k} \sim 10^{122}$ (i.e. $\rho_{\gamma} \sim 10^{-10}$ GeV cm$^{-3}$ and $\rho_{m} \sim 10^{-5}$ GeV cm$^{-3}$) 
correspond to $n \sim 10^{122}$ and fluctuations $\delta a \sim l \sim 10^{28}$ cm. In such a description, the observable part of the universe 
appears as a gigantic fluctuation.

Under the assumption that a coefficient function that connects the root-mean-square deviations of scale factor and position of a test particle in 
coordinate space, as well as a coefficient function that connects the root-mean-square deviations of their conjugate momenta, can be taken in the 
form of the normal frequency functions, the generalized uncertainty principle proposed previously by a number of authors can be reproduced.
The uncertainty relation (\ref{16}) can be rewritten in an equivalent form (\ref{123}) relating the standard 
deviations of position and momentum of a test particle, so that the right-hand side of the inequality will include the effects of gravity explicitly.
This term generates the modified Heisenberg algebra
\begin{equation*}
[x,p] = \hbar \frac{\Delta x}{\Delta a} \frac{\Delta p}{\Delta \pi} = 
\hbar \left(1 +  \frac{1}{2 L_{x}^{2}} (\Delta x)^{2} + \frac{L_{p}^{2}}{2} \left(\frac{\Delta p}{\hbar} \right)^{2} + \ldots \right).
\end{equation*}
The implementation of this algebra results in the appearance of additional terms in the Hamiltonian of a test particle, which will describe 
gravitational effects. The case, for which the fluctuation $(\Delta x)^{2}$ is neglected, is considered in Refs.~\cite{Kem,Bra,Das,Ber}
(see also the review article \cite{Cha}).

\end{document}